\documentclass[12pt, preprintnumbers,amsmath,amssymb]{revtex4}
\usepackage{latexsym}
\usepackage{graphics}
\usepackage{epsfig}
\usepackage{dcolumn}
\def\br{\begin{eqnarray}}
\def\er{\end{eqnarray}}
\def\be{\begin{equation}}
\def\ee{\end{equation}}

\def\L{\Lambda}

\def\({\left(}
\def\){\right)}
\def\<{\left\langle}
\def\>{\right\rangle}

\begin{document}
%\twocolumn[\hsize\textwidth\columnwidth\hsize\csname %%% TWO COLUMN
%@twocolumnfalse\endcsname                            %%% TWO COLUMN
%
%\draft
%

\title{Walking Behavior in Technicolored GUTs}
\author{A. Doff$^{a}$} 
\affiliation{$^a${\small Universidade  Tecnol\'ogica  Federal do Paran\'a - UTFPR - COMAT, Via do Conhecimento Km 01, 85503-390, Pato  Branco, PR, Brazil}}

%%%%
\date{\today}
\begin{abstract} 
There exist two ways to obtain walk behavior:
assuming a large number of technifermions in the fundamental
representation of the technicolor (TC) gauge group,
or a small number of technifermions, assuming that these
fermions are in higher-dimensional representations of the
TC group. We propose a scheme to obtain the walking behavior
based on technicolored GUTs (TGUTs), where elementary
scalars with the TC degree of freedom may remain
in the theory after the GUT symmetry breaking.
\end{abstract}
%\pacs{12.60.Cn, 12.60.Rc, 11.30.Na} 
\maketitle
%\vskip 0.2cm]
\newpage 
\section{Introduction}
%\label{intro}
The nature of the Higgs boson is one of the most important
problems in particle physics, and there are many questions
that may be answered in the near future by the LHC experiments.
Is the Higgs boson, if it exists at all, elementary
or composite? What are the symmetries behind the Higgs
mechanism? The possibility that the Higgs boson is a composite
state instead of an elementary one is more akin to
the phenomenon of spontaneous symmetry breaking that
originated from the effective Ginzburg\,–\,Landau Lagrangian,
which can be derived from the microscopic BCS theory of
superconductivity describing the electron–hole interaction
(or the composite state in our case). This dynamical origin of
spontaneous symmetry breaking has been discussed with the
use of many models, the most popular one being the technicolor
(TC) model\cite{lane}. 
\par Unfortunately we do not know the dynamics that forms
the scalar bound state, which should play the role of the
Higgs boson in the standard-model symmetry breaking, and
no phenomenologically satisfactory model along this line
has been derived up to now. Most of the models for the spontaneous
symmetry breaking of the standard model based on the composite Higgs boson 
system depend on specific assumptions
about the particle content and consequently on
the dynamics responsible for the bound-state formation\cite{hs}. In theories
based in TC the Higgs boson is a composite of
the so-called technifermions, and to some extent any model
in which the Higgs boson is not an elementary field follows
more or less the same ideas as in technicolor models. The
beautiful characteristics of TC as well as its problems were
clearly listed recently by Lane\cite{lane,hs}. Most of the technicolor 
problems may be related to the dynamics of the theory as described in Ref.\cite{lane}.
\par  Although technicolor is a non-Abelian gauge theory it
is not necessarily similar to QCD, and if we cannot even
say that QCD is fully understood up to now, it is perfectly
reasonable to realize the enormous work that is needed to
abstract from the fermionic spectrum the underlying technicolor
dynamics. The many attempts to build a realistic
model of dynamically generated fermion masses are reviewed
in Ref.\cite{lane,hs}. Most of the work in this area tries to find the
TC dynamics dealing with the particle content of the theory
in order to obtain a technifermion self-energy that does not
lead to phenomenological problems as in the scheme known
as walking technicolor\cite{walk}. The idea of this scheme is quite simple. 
First, we can remember that the\, expression for the\, TC self-energy is proportional to $ \Sigma (p^2)_{{}_{TC}} \propto { (\langle \bar{\psi} \psi\rangle_{{}_{TC}}}/{p^2})( p^2/\L^2_{{}_{TC}})^{\gamma_{{}_{TC}}}$, where $\langle \bar{\psi}\psi\rangle_{{}_{TC}}$ is the TC condensate and $\gamma_{{}_{TC}}$ its anomalous dimension. Secondly, depending
on the behavior of the anomalous dimension we obtain different behaviors for $\Sigma (p^2)_{{}_{TC}}$. 
\par In principle, we could deal with many different models, varying fermion representations and particle content, finding different expressions for $ \Sigma (p^2)_{{}_{TC}}$ and testing them phenomenologically, i.e. obtaining the fermion mass spectra without any conflict with experiment. The anomalous dimension  $\gamma_{{}_{TC}} = 1$ can be obtained in the extreme limit of
a walking technicolor dynamics, and a large anomalous dimension
may solve many problems in TC models\cite{an}.
\par Usually the walking behavior is obtained assuming a
large number of technifermions, $N_{TF} \sim 4N_{TC}$, if technifermions
are in the fundamental representation of TC gauge group\cite{walk}. Moreover, recently Sannino et al.
showed that it is possible to obtain the walking behavior for a small number of technifermions if these are in higher dimensional
representations of the TC gauge group\cite{sannino}. 
\par In this work we propose a scheme that combines these
two approaches to obtain the walking behavior based on
technicolored GUTs, in the sense that we have a reduced
number of technifermions in the fundamental representation
of TC group, and technicolored Higgs bosons in higher dimensional
representations of the TC group. The advantage
of this approach in relation the other proposes to generate the
walking behavior is that in this case the dynamics associated
to the TC would be similar to the QCD until an energy scale
of order $O(10^{13}GeV)$.
\par This paper is composed as follows: in Sect. 2 we discuss
the modification of the TC beta function ($\beta_{{}_{TC}}$) in the presence
of scalars that carry technicolor degrees of freedom.
After that, in Sect. 3, we propose a "toy model" for a technicolored
GUT (TGUT) to illustrate how we can obtain the
walking behavior. In Sect. 4 we draw our conclusions.

\section{The TC Beta Function in the Presence of Scalars}
%\label{sec:2}

\par For asymptotically free gauge theories, with fermions in the
fundamental representation, the $\beta$ function at $g^5$ order (two-loop) can by written in the following form\cite{jones1} 
\be 
\beta(g) = -b_{0}\frac{g^3}{16\pi^2} + b_{1}\frac{g^5}{(16\pi^2)^2}
\ee  
\noindent where we defined 
\br 
&&\!\!\!\!\!\! b_{0}= \frac{11}{3}C_2(G) - \frac{4}{3}\!\!\!\sum_{{}_{\rm fermions}}\!\!T(R)\nonumber \\
&&\!\!\!\!\!\! b_{1}= -\frac{34}{3}C^{2}_{2}(G) + \frac{20}{3}C_{2}(G)\!\!\!\!\sum_{{}_{\rm fermions}}\!\!\!\!T(R) + 4C_2(R)\!\!\!\!\sum_{{}_{\rm fermions}}\!\!\!\!T(R). \nonumber \\ 
\er 
\noindent For $SU(N)$ and fermions in the fundamental representation these coefficients assume the  form  $C_{2}(G) = N$, $T(R)= \frac{1}{2}$, and $C_2(R) = \frac{N^2 - 1}{2N}$. 
\par  The $\beta$ function in the presence of scalars at two loop approximation  was  determinated by Jones\cite{jones2} and the form of the coefficients  $b_{0}$ and $ b_{1}$ translates to 
\br 
b_{0} &=& \frac{11}{3}C_2(G) - \frac{4}{3}\!\!\!\sum_{{}_{\rm fermions}}\!\!T(R) - \frac{1}{3}\!\!\!\sum_{{}_{\rm scalars}}\!\!T(S)\nonumber \\
b_{1} &=& -\frac{34}{3}C^{2}_{2}(G) + \frac{20}{3}C_{2}(G)\!\!\!\sum_{{}_{\rm fermions}}\!\!T(R)  + \nonumber \\
      &+& 4C_2(R)\!\!\!\sum_{{}_{\rm fermions}}\!\!T(R) + 4\!\!\!\sum_{{}_{\rm scalars}}\!\!C_{2}(S)T(S) \nonumber \\
      &+& \frac{2}{3}\!\!\!\sum_{{}_{\rm scalars}}\!\!C_{2}(G)T(S). 
\label{betascalar}
\er 
\par In the notation used in the equations described above the
indices $R$ and $S$ respectively denote fermions and scalars
in the fundamental representation; in this case the form of
the coefficients $C_2(S)$ and $T(S)$ is the same for the fermions.
However, we can determine the form of the coefficients
$C_2(S)$ and $T(S)$ for a generic representation r considering
the relations
\br 
&& (C_{2}(r_{1}) + C_{2}(r_{2}))d(r_{1})d(r_{2}) = \sum C_{2}(r_{i})d(r_{i}), \nonumber \\ 
&& T(r)d(G) = d(r)C_{2}(r)
\label{const}
\er 
\noindent where $d(r_{i})$ is the dimension of the representations $r_{i}$ with $i=1,2$. 
\par  We can determine the values of the coefficients $T(R,S)$ and $C_{2}(k)$, where $k = G,R,S$,  to some TC group.
 Moreover, it is interesting to consider the case where $N_{TC} = 3$,
because in the next section we will build a model based on
this specific group. Therefore, after we consider the relations
shown in (4), we obtain
\br 
&&C_{2}(S = 3) = \frac{4}{3}\,\,\,,\,\,\,T(S = 3) = \frac{1}{2}\nonumber \\ 
&&C_{2}(S = 6) = \frac{10}{3}\,\,,\,\,\,T(S = 6) = \frac{5}{2} \nonumber \\ 
&&C_{2}(S = 8) = 3\,\,\,\,\,,\,\,\,\,\,T(S = 8) = 3.
\er 
\noindent The coefficients $b_{0}$ and $ b_{1}$ described by (3) can be written
\br 
 && b_{0}=  11 - \frac{2}{3}n_{f} - \frac{1}{6}n_{{}_{S}}(3) - \frac{5}{6} n_{{}_{S}}(6) -  n_{{}_{S}}(8)\nonumber \\
 && b_{1}=  -102 + \frac{38}{3}n_{f} + \frac{11}{3} n_{{}_{S}}(3) + \frac{115}{3} n_{{}_{S}}(6) + 42n_{{}_{S}}(8), \nonumber \\
\er
\noindent where $n_f$ is the number of Dirac fermions  needed to generate the walking behavior,  and $n_s$ is the number of introduced complex scalars.
\par In the introductory section we present the expression for
the TC self-energy, which for convenience we repeat:
\be 
 \Sigma (p^2)_{{}_{TC}}\,\, \approx\,\, \frac{\langle \bar{\psi} \psi\rangle_{{}_{TC}}}{p^2}\left( \frac{p^2}{\Lambda^2_{{}_{TC}}}\right)^{\gamma_{{}_{TC}}} = \,\,\, \frac{\Lambda^{3}_{{}_{TC}}}{p^2}\left( \frac{p^2}{\Lambda^2_{{}_{TC}}}\right)^{\gamma_{{}_{TC}}}.
\ee 
\noindent As we argue, the form for the TC self-energy will depend on
the value assumed by the anomalous dimension $\gamma_{{}_{TC}}$, and the
walking behavior $\gamma_{{}_{TC}} \approx 1$ is obtained in conventional models
assuming $N_{TF} \sim 4N_{TC}$. For example, to $SU(N_{TC} = 3)$,
the number of technifermions necessary to generate this behavior
is $N_{TF} \sim 12 $. Moreover, in this scenario we have a
great proliferation of pseudo-Goldstone bosons (PGBs) and in general large contributions to the parameters of precision
S, T and U due to the high number of technifermions introduced.
  
\par In the scenario proposed in this work the number of technifermions
needed to generate the walking behavior would
be much less. Because if we consider (1) and (6), the
walking behavior $\beta(g)_{TC} \approx 0$, for  the  case $SU(N_{TC} = 3)$, is
obtained approximately by the following number of fermions\cite{comment}:
\be
N_{TF} \sim 12 - \frac{2}{3}n_{{}_{S}}(3) - \frac{7}{3}n_{{}_{S}}(6) - 3n_{{}_{S}}(8). 
\ee 
\noindent In the next section we will propose a "toy model" for a technicolored
GUT (TGUT) to illustrated our ideas, and in this
specific case the gauge group attributed to TC is precisely $SU(N_{TC} = 3)$. 

\section{A Toy Model for a  Technicolored GUT}

%\label{sec:3}

\par In this section we propose a "toy model" for a technicolored
GUT (TGUT) based on the group $SO(10)$ in order to illustrate
our approach. We shall begin by presenting the usual
fermionic content attributed to the $SO(10)$ model,
\br 
{\bf{16}}^T(GUT) = ( \psi^T, \phi^T)_{L}\nonumber \\
\er 
\noindent where we denote
\br
&&\psi^T =  (u^a_{1}...u^a_{3}\, ,\nu^{a}, d^a_{1}... d^a_{3} \,, l^{a}) \nonumber \\
&&\phi^T = (d^{[a]c}_{1}...d^{[a]c}_{3}\,,  l^{[a]c} \,, -u^{[a]c}_{1}... -u^{[a]c}_{3}\,, -\nu^{[a]c}),\nonumber \\
\label{so10}
\er
\noindent and in this expression  $(a,i=1..3)$ are respectively
the family and color indices. To exemplify our purpose of
obtaining the walking behavior we shall consider a "toy
model" that just contains a complete generation of leptons $\psi^{T}_{l} = (l^a, \nu^a)$
and techniquarks $\psi^{T}_{Q} = (U^a, D^a)$. In this
case we can consider a model with the fermionic content exhibited
in (9) and just make the change $(u,d) \rightarrow (U,D)$. 
\br 
{\bf{16}}^T(TGUT) = ( \Psi^T, \Phi^T)_{L}\nonumber \\
\er 
\noindent where
\br
&&\Psi^T =  (U^a_{1}...U^a_{3}\, ,\nu^{a}, D^a_{1}... D^a_{3} \,, l^{a}) \nonumber \\
&&\Phi^T = (D^{[a]c}_{1}...D^{[a]c}_{3}\,,  l^{[a]c} \,, -U^{[a]c}_{1}... -U^{[a]c}_{3}\,, -\nu^{[a]c}).\nonumber \\
\er
\noindent In the above expression we can identify $(U,D)$ as the usual
techniquarks and the index  $i=1..3$ labels the technicolor.
The spontaneous symmetry breaking of the model to $SU(3)_{TC}\times SU(2)_{L}\times U(1)_{Y}$
can be promoted assuming a convenient set of scalar fields that develop nonzero vacuum
expectation values (VeVs). At the low-energy scale the technicolor
theory breaks the electroweak symmetry dynamically by the formation of a condensate
$\langle \bar{Q}Q\rangle  \sim \Lambda^{3}_{{}_{TC}}$.

\par  The symmetry breaking pattern $SO(10) \rightarrow SU(3)_{TC}\times SU(2)_{L}\times U(1)_{Y}$ can be obtained as described in Ref.\cite{kuzmin}
\be 
SO(10) \stackrel{54,16}{\rightarrow} SU(3)_{TC}\times SU(2)_{L}\times U(1)_{Y}.
\ee
 \noindent For this breaking pattern the VeVs attributed to the Higgs
fields {\bf 54} and {\bf 16} are chosen to be of the same order, $V_{54} \approx V_{16} \sim \Lambda_{TGUT}$, and we do not have any other intermediate
stage. The physical spectrum of the Higgs fields can be extracted from\cite{babu} and in Table I we list these fields and
these respective masses in terms of the  $SU(3)_{TC} \times SU(2)_{L} \times U(1)_{Y}$ decomposition.
 
\par In $(14)$ we show the usual expression for the technicolor
anomalous dimension, where only the usual technifermions
were included. The plot of the behavior of this anomalous
dimension, $\gamma_{{{}_{TC}}_{(A)}}$ to  $N_{{}_{TC}} = 3$
as a function of $\Lambda$, is depicted in Fig. I(A):
\br
\gamma_{{{}_{TC}}_{(A)}} &=& \frac{3(N^{2}_{{}_{TC}} - 1)}{4\pi N_{{}_{TC}}}\alpha_{{}_{TC}}(\Lambda^2) \nonumber \\
                 &=& \frac{3(N^2_{{}_{TC}} - 1)}{4\pi N_{{}_{TC}}}\frac{\alpha_{{}_{TC}}(\Lambda^2_{{}_{TC}})}{1 + (\frac{b_{0}}{2\pi} + \frac{b_{1}}{8\pi^2})\ln\left(\frac{\Lambda^2}{\Lambda^2_{{}_{TC}}}\right)},
\er 
\noindent where the coefficients  $b_{0}$ and $b_{1}$ are described in $(2)$. However,
at high energies the technicolored Higgs bosons responsible for the breaking of the TGUT 
group contribute to the running of the technicolor coupling and the expression
for the technicolor anomalous dimension for $N_{{}_{TC}} = 3$ can
be written as
\br
\gamma_{{{}_{TC}}_{(B)}} &=& \frac{3(N^{2}_{{}_{TC}} - 1)}{4\pi N_{{}_{TC}}}\alpha_{{}_{TC}}(\Lambda^2) \nonumber \\
                 &=& \frac{3(N^2_{{}_{TC}} - 1)}{4\pi N_{{}_{TC}}}\frac{\alpha_{{}_{TC}}(\Lambda^2_{{}_{TC}})}{1 + (\frac{b_{0}}{2\pi} + \frac{b_{1}}{8\pi^2})\ln\left(\frac{\Lambda^2}{\Lambda^2_{{}_{TC}}}\right)}
\er 
\noindent in this case, the coefficients $b_{0}$ and $b_{1}$  are described by (6).
The plot of the behavior of this anomalous dimension, $\gamma_{{{}_{TC}}_{(B)}}$,  as a function of $\Lambda$ is depicted in Fig. I(B).

\begin{figure}[t]
\resizebox{0.8\textwidth}{!}{%
  \includegraphics{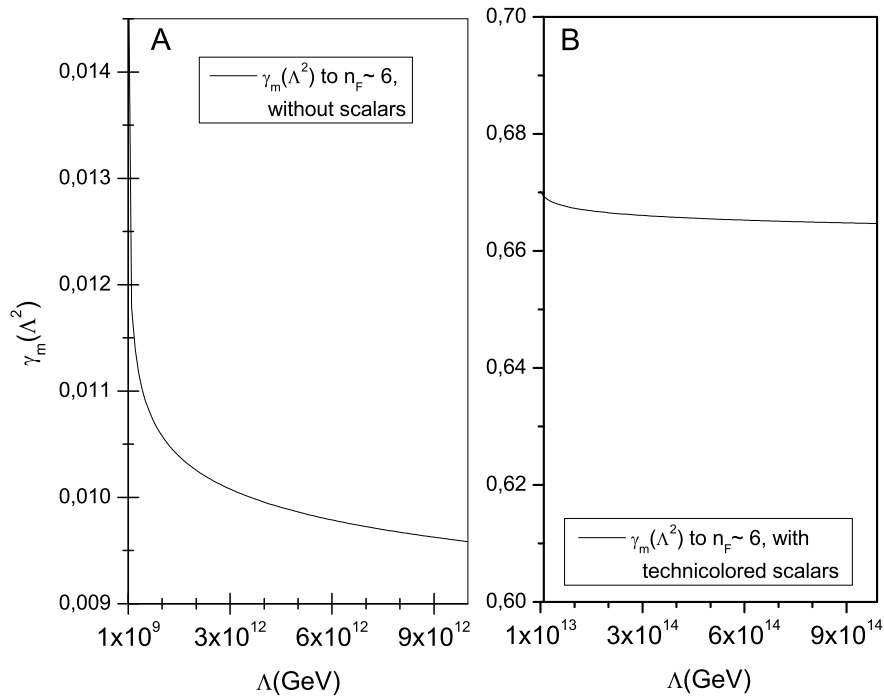}
}
%\vspace{5cm}       % Give the correct figure height in cm
\caption{Evolution of the  anomalous dimension $\gamma_{{}_{TC}}$ as function of the number of technifermions ($N_{TF}$) and technicoloured heavy scalars ($n_{{}_{S}}(r)$, to $r = 3,6,8$). To achieve  this plot  we assumed the following values, $\Lambda_{TC}  =  300 GeV$, and $\alpha_{{}_{TC}}(\Lambda^2_{{}_{TC}})  \approx  \pi/3 \approx 1$.}
\label{fig:1}       % Give a unique label
\end{figure}
\noindent From this graph it is possible  verify that the number of technifermions  necessary to obtain the walking behavior is approximately $N_{TF} \sim 6$. Then, for energies of the order $O(10^{13}GeV)$, the TC self-energy presents a more weaker dependence with momentum 
\be 
\Sigma (p^2)_{{}_{TC}} \sim \,\,\, \frac{\Lambda^{2}_{{}_{TC}}}{p}\left( \frac{p}{\Lambda_{{}_{TC}}}\right)^{\frac{1}{5}}.
\ee 
\begin{table}[t]
\caption{We present the masses obtained by technicolored Higgs
bosons. The constants $\alpha_{1}$ and $\alpha_{2}$  are associated to the quadrilinear
Higgs coupling involving, respectively, the {\bf 54}[$\chi,\Phi$] and {\bf 16}[$\eta$] Higgs fields.}
\label{tab:1}       % Give a unique label
\begin{tabular}{lll}
\hline\noalign{\smallskip}
 Technicolored   &  Decomposition under     &  Mass      \\ 
 Higgs Bosons    &   $SU(3)_{TC} \times SU(2)_{L} \times U(1)_{Y}$   &  \\
\noalign{\smallskip}\hline\noalign{\smallskip}
       $\chi$        &    (8,1,0)     &     $\sim  \alpha_{1}\Lambda_{TGUT} $      \\ 
       $\Phi$        &    (6,1,-2/3)  &     $\sim  \alpha_{1}\Lambda_{TGUT} $      \\ 
       $\eta$        &    (3,1,1/3)   &     $\sim  \alpha_{2}\Lambda_{TGUT}$      \\  
\noalign{\smallskip}\hline
\end{tabular}
% Or use
%\vspace*{5cm}  % with the correct table height
\end{table}
\par This behavior is reached just at the scale of energy where
the new degrees of freedom associated to the technicolored
Higgs bosons appear. The ideas that we presented in this paper
are not new. A long time ago the authors of\cite{colorhiggs1,colorhiggs2} 
suggested that heavy colored Higgs scalars, which are present
in theories such as $SU(5)_{GG}$ should be included in the evolution
of the $\beta$ function  of QCD at high energies. As the
authors of\cite{colorhiggs2} argue, if the masses of these Higgs bosons
are of the order  $\alpha\Lambda_{GUT}$, where $\alpha \sim O(10^{-2})$,
these degrees of freedom would then modify to high energies the classic
value obtained for $\sin\theta_{W}$ in the  $SU(5)_{GG}$
 because at two loop order the running coupling constants of the Standard Model 
 $\alpha_{3}, \alpha_{2}$ and $\alpha_{1}$ mix\cite{colorhiggs1}, where 
\be 
\frac{d\alpha_{i}^{-1}}{dt} = \frac{\beta_{0}^{i}}{4\pi} + \sum^{\!{}_{3}}_{{}_{\!\!j=1}}\beta_{1}^{ij}\frac{\alpha_{j}}{(4\pi)^2}
\ee 
\noindent and in the expression above $i, j = 1..3$ label the $U(1)$, \,$SU(2)$ and $SU(3)$ gauge groups.
\par  In this work we just extended some of these ideas to TC
models, showing that it is possible to obtain the walking behavior
with a reduced number of technifermions in the fundamental
representation of the TC group if we have technicolored
scalar bosons resulting from a TGUT.

\section{Conclusions}
%\label{sec:concl}

\par In summary, in this paper we presented a possible way to
generate the walking behavior in TC models using scalar
matter besides ordinary fermionic matter. In the scheme proposed
we get walking with a reduced number of technifermions
in the fundamental representation of the TC group and
the TC dynamics is similar to QCD until an energy scale of
order $O(10^{13}GeV)$. 
\par At the energy scale where fermionic mass would be generated,
near the TGUT scale, the behavior of the technicolor  $\beta$ function
would be modified by the presence of the degrees of freedom associated
to the heavy technicolored Higgs bosons that result from the TGUT breaking, 
leading to the walking behavior.
\par The unification group considered in this work is not realistic.
It was used only to illustrate our proposal to obtain
walking, though we believe that these ideas can be applied to
more realistic models. As an example of a future proposal,
which includes the degrees of freedom of color, we want
to propose a model along the lines of the Farhi\,–\,Susskind
model\cite{farhi}.

\section*{Acknowledgments}

I would like to thank A. A. Natale for discussions and the Conselho Nacional de Desenvolvimento  Cient\'{\i}fico e Tecnol\'ogico (CNPQ) and Funda\c c\~ao de Apoio ao Desenvolvimento Cient\'{\i}fico e Tecnol\'ogico do Paran\'a (Funda\c c\~ao Arauc\'aria) by  financial support.

%
% BibTeX users please use
% \bibliographystyle{}
% \bibliography{}

\begin{thebibliography}{12}
\bibitem{comment} Notice that this is a rough approximation to  critical number of technifermions  necessary to get the walking behavior after we introduce the set of the technicolored scalars  listed in Table I.
\end{thebibliography}

\end{document}